







\message{ Assuming 8.5" x 11" paper }    

\magnification=\magstep1	          

\raggedbottom

\parskip=9pt

%

\def\singlespace{\baselineskip=12pt}      
\def\sesquispace{\baselineskip=16pt}      



\font\openface=msbm10 at10pt
\def\Minkowski     {{\hbox{\openface M}}}
\def\less{\backslash}		
\def\Integers      {{\hbox{\openface Z}}}
\def\implies{\Rightarrow}
\def\eps{\varepsilon}
\def\SetOf#1#2{\left\{ #1  \,|\, #2 \right\} }

\def\author#1 {\medskip\centerline{\it #1}\bigskip}
\def\address#1{\centerline{\it #1}\smallskip}
\def\furtheraddress#1{\centerline{\it and}\smallskip\centerline{\it #1}\smallskip}
\def\email#1{\smallskip\centerline{\it address for email: #1}} 

\def\AbstractBegins
{
 \singlespace                                        
 \bigskip\leftskip=1.5truecm\rightskip=1.5truecm     
 \centerline{\bf Abstract}
 \smallskip
 \noindent	
 } 
\def\AbstractEnds
{
 \bigskip\leftskip=0truecm\rightskip=0truecm       
 }

\def\section #1 {\bigskip\noindent{\headingfont #1 }\par\nobreak\smallskip\noindent}

\font\titlefont=cmb10 scaled\magstep2 
\font\headingfont=cmb10 at 12pt
\font\subheadfont=cmssi10 scaled\magstep1 
\font\csmc=cmcsc10  
\def\NOTATION{\noindent {\csmc Notation \ }} 
\def\DEFINITION{\noindent {\csmc Definition \ }} 
\def\REMARK{\noindent {\csmc Remark \ }}
\def\THEOREM{\noindent {\csmc Theorem \ }}
\def\subsection #1 {\medskip\noindent{\subheadfont #1 }\par\nobreak\smallskip\noindent}
\def\ReferencesBegin
{
 \singlespace					   
 \vskip 0.5truein
 \centerline           {\bf References}
 \par\nobreak
 \medskip
 \noindent
 \parindent=2pt
 \parskip=6pt			
 }
\def\reference{\hangindent=1pc\hangafter=1} 

\def\ref{\reference}

\def\sepref{\parskip=8pt \par \hangindent=1pc\hangafter=0}
\def\journaldata#1#2#3#4{{\it #1\/}\phantom{--}{\bf #2$\,$:} $\!$#3 (#4)}
\def\linebreak{\hfil\break}
\def\lbr{\linebreak}
\def\eprint#1{{\tt #1}}
\def\arxiv#1{\hbox{\tt http://arXiv.org/abs/#1}}






\def\Mink{\Minkowski}
\font\german=eufm10 at 10pt
\def\Buchstabe#1{{\hbox{\german #1}}}
\def\AA{\Buchstabe{A}}

\def\PROOF{\noindent \quad{\csmc Proof \ }}
\def\LEMMA{\smallskip\noindent {\csmc Lemma }}

\def\sepref{\parskip=8pt \par \hangindent=1pc\hangafter=0}



\phantom{}





\sesquispace
\centerline{{\titlefont Symmetry-breaking and zero-one laws}}

\bigskip


\singlespace			        

\author{Fay Dowker}
\address{Perimeter Institute, 31 Caroline Street North, Waterloo ON, N2L 2Y5 Canada}
\furtheraddress{Blackett Laboratory, Imperial College, Prince Consort Road, London SW7 2AZ, UK}
\email{f.dowker@imperial.ac.uk}

\medskip
\centerline{\bf and}
\smallskip

\author{Rafael D. Sorkin}
\address
 {Perimeter Institute, 31 Caroline Street North, Waterloo ON, N2L 2Y5 Canada}
\furtheraddress
 {Raman Research Institute, C.V. Raman Avenue, Sadashivanagar, Bangalore -- 560 080 India}
\furtheraddress
 {Department of Physics, Syracuse University, Syracuse, NY 13244-1130, U.S.A.}
\email{rsorkin@perimeterinstitute.ca}

\AbstractBegins                              
We offer further evidence that discreteness of the sort inherent in a
causal set cannot, in and of itself, serve to break Poincar{\'e}
invariance.  In particular we prove that a Poisson sprinkling of
Minkowski spacetime cannot 
endow spacetime with a distinguished spatial or temporal orientation,
or with a distinguished lattice of spacetime points,
or with a distinguished lattice of timelike directions
(corresponding respectively to breakings of reflection-invariance,
translation-invariance, and Lorentz invariance).
Along the way we provide a proof from first principles of the zero-one
law on which our new arguments are based.
\bigskip
\noindent {\it Keywords and phrases}:  discreteness, symmetry breaking, zero-one law, Poisson process, causal set, quantum gravity
\AbstractEnds                                

\bigskip


\sesquispace
\vskip -10pt

\section{Introduction}                       
Will a discrete structure prove to be the kinematical basis of quantum
gravity and if so should we expect it to preserve the known symmetries
of Minkowski spacetime, at least quasi-locally?  One strand of thought
has tended to answer these questions with ``yes'' followed by ``no'',
and has held out effects like modified dispersion relations for
electromagnetic waves as promising candidates for a phenomenology of
spatiotemporal discreteness.  In contrast we have maintained in earlier
work that the type of discreteness inherent in a causal set
cannot, in and of itself, serve to break Poincar{\'e} invariance.  In
[1] we offered informal arguments to this effect, and then in
[2] it was proved rigorously that a ``sprinkling'' of Minkowski
spacetime induced by a Poisson process can determine a rest frame only
with zero probability.

This theorem, however, left open the possibility that a sprinkling, even
if it could not remove all the symmetry of flat spacetime, could
nevertheless cut it down to a proper subgroup $H$ of the Poincar{\'e} group
$G$.  In this paper we will address that possibility, and provide
further evidence against it, proving in particular that 
a Poisson sprinkling of Minkowski spacetime 
cannot induce an ``arrow of time'' or a ``chirality'',
that it cannot break translation-symmetry by endowing spacetime with 
a distinguished lattice of points,
and that it cannot break Lorentz symmetry by endowing spacetime with 
a distinguished ``lattice'' of timelike directions.
%
More generally we
conjecture that a sprinkling will almost surely preserve the full
group $G$, and we explain how one can potentially corroborate this
expectation in any particular case (i.e. for any putative pattern of
symmetry breaking) by 
combining the methods of this paper with those of [2].

Our new method herein will rely on a certain ``zero-one law'' that governs
invariant events in the theory of Poisson processes.  To make the paper
more self-contained, and also to provide a result of the requisite
strength, we have chosen to prove the main zero-one theorem starting
from nothing but general facts about probability measures.  The
resulting demonstration seems to us to be as simple as possible, and we
hope that along with the proof per se, some of the definitions and
lemmas that lead up to the main theorem will prove to be of independent
interest.

After presenting and proving these lemmas in the next section of the
paper, we prove the main theorem and then show how to apply it to
exclude symmetry-breaking, first in important special cases, and then
conjecturally in the general case.
We also take the opportunity to reply, in an Appendix, to some recent
criticism of the theorems proven in~[2]. 

For further background on these questions we refer the reader to
[1] and [2].

\section{Preparing to prove a zero-one law}  
In the next section, we will prove a ``zero-one law'' about Poisson
processes, from which will follow the desired theorems on
symmetry-preservation in many, if not all, cases of interest.  In fact,
a version of this result can be found in [3], but that theorem would
not let us rule out certain important cases of symmetry-breaking. For
example it would not let us exclude that a sprinkling might break the
group of all translations down to a discrete subgroup, as happens for
example when a liquid crystallizes.
For this reason, we have decided to demonstrate ab initio the zero-one
law we will be appealing to herein.  We hope also that our development
will help to clarify how and why such laws arise.
In preparation, let's first review some definitions and known results
from [2] and [4].

Let $\mu$ be the measure that, mathematically speaking, defines our
sprinkling process, which we take to be a Poisson process in $\Mink^n$,
the Minkowski space of dimension $n$.
An individual sprinkling in $\Minkowski^n$ is almost surely a locally
finite subset of $\Minkowski^n$.  The space of all such subsets, which
we will denote by $\Omega$, is the {\it\/sample space\/} of the Poisson
process.  A measurable subset of $\Omega$ will be called an
{\it\/event\/}, as is customary for stochastic processes.  The set of
all events forms a $\sigma$-algebra that we will call the
{\it\/event-algebra\/} $\AA$.  

The concept of a {\it\/bounded event\/} will important for our proof.
By definition such an event will be 
one that 
pertains to
a bounded (say compact) subset of $\Mink^n$, 
by which we mean more precisely the following.  
Let $\omega\in\Omega$ be any sprinkling, and $B$ a subset of $\Mink^n$.
We say that an event $A$ 
is ``an event within $B$''
(or is ``supported within $B$'') 
if
in order to know whether $\omega\in A$ it suffices to know the subset, 
$\omega\cap{B}$, 
of sprinkled points that fall within $B$.  
For example the event, ``There are more than 5 sprinkled points in $B$'', is an event within $B$.
We call an event bounded iff it is an event within $B$ 
for some bounded spacetime region $B$. 

We will write $\AA_0$ for the set of all bounded events.  It is not
a $\sigma$-algebra, but it is still a Boolean algebra, meaning it is
closed under the operations of Boolean sum and Boolean product, as
defined below.  
Equivalently it is closed under union, intersection, and set-difference.

It will important for our proof that every event $A\in\AA$ can be built
up as a (countable) logical combination of bounded events.  
Formally, this says that the full event-algebra $\AA$ is generated qua
$\sigma$-algebra by $\AA_0$. 
(This basic fact about Poisson processes results directly from the 
way in which they are defined [5] [3].)
We claim (and will shortly prove) that as a consequence, 
every event in $\AA$ is 
in a well-defined sense 
a limit of bounded events.

Before turning to the proof, we need to establish a few more
definitions and some notation and lemmas.  Most of the lemmas are either
well known or easy to prove, but we include them for completeness, and
because some of our definitions are not quite the usual ones.

\NOTATION
Let $A$ and $B$ be events.  
Their {\it\/boolean sum\/}, $A+B$, is their ``symmetric difference'', $(A\cup{B})\less(A\cap{B})$.
Their {\it\/boolean product\/}, $AB$, is their intersection, $A\cap{B}$.

\noindent
This little-used but convenient notation exhibits explicitly that the
events form an algebra over $\Integers_2$, with identity $1$ equal to
the event $\Omega$.  The complement of an event $A$ can thus be written
as $1+A$.

\DEFINITION (``distance'' between two events): $d(A,B)=\mu(A+B)$

\DEFINITION Let $A, A_1,A_2,A_3\dots$ be events in $\AA$. Then $A_k\to{A}$ means that $d(A_k,A)\to0$.  

\noindent We will also say in this situation that $A$ is a limit of the $A_k$.

\noindent The next two lemmas will verify  the
triangle-inequality for $d$.  The latter is not technically a metric,
however, because $d(A,B)=0$ does not imply that $A=B$.

\LEMMA 1. \  $\mu(A+B) \le \mu(A) + \mu(B)$

\PROOF $A+B\subseteq A\cup B \implies \mu(A+B) \le \mu(A\cup B) \le \mu(A) + \mu(B)$.

\LEMMA 2 (triangle inequality). \   $d(A,C)\le d(A,B) + d(B,C)$

\PROOF $A+C = (A+B) + (B+C)$ because $B+B=2B=0$.  
 Hence, in light of the previous lemma, $\mu(A+C) \le \mu(A+B) + \mu(B+C)$.

\LEMMA 3. \   $|\mu(A)-\mu(B)| \le \mu(A+B)$

\PROOF  A Venn diagram makes this clear.  More computationally, we have,
since the measure $\mu$ is additive, 
$\mu(A) = \mu(A\less B) + \mu(AB)$,
and similarly
$\mu(B) = \mu(B\less A) + \mu(AB)$,
whence
$\mu(A)-\mu(B)=\mu(A\less B)-\mu(B\less A) \le \mu(A\less B)+\mu(B\less A)=\mu(A+B)$,
and similarly
$\mu(B)-\mu(A)\le\mu(A+B)$.

\noindent
From this last lemma follows immediately the continuity of $\mu$ with
respect to $d$, as well as that of addition and multiplication.

\LEMMA 4. \   $A_j\to A \implies \mu(A_j)\to\mu(A)$

\LEMMA 5. \   $A_j\to A$ and $B_j\to B$  $\implies$ $A_j B_j \to AB$ and $A_j + B_j \to A+B$

\noindent (In other words limit preserves boolean sum and product.)

\PROOF  First notice that 
$A_jB_j + AB = A_j(B_j+B) + (A_j+A)B$, and that
$A_j(B_j+B)\subseteq (B_j+B)$, while
$(A_j+A)B\subseteq A_j+A$. 
Therefore
$d(A_j B_j,AB)=\mu(A_j B_j+AB) \le \mu(B_j+B)+\mu(A_j+A)=d(B_j,B)+d(A_j,A)\to 0$.
The proof for $A+B$ is similar but simpler.  Start with the trivial equation,
$(A_j+B_j)+(A+B)=(A_j+A)+(B_j+B)$
and apply $\mu$ to both sides.  The result is
$d(A_j+B_j,A+B) = \mu[(A_j+A)+(B_j+B)] \le \mu(A_j+A) + \mu(B_j+B) = d(A_j,A) + d(B_j,B) \to 0$

\REMARK  We could prove in the same way that limit preserves complementation:
$A_j\to A$ $\implies$ $1+A_j\to 1+A$,
but it follows already from the lemma.

\noindent
The next lemma holds for any Boolean algebra of events and the
$\sigma$-algebra it generates.  
\LEMMA 6. Every event in $\AA$ is the limit of a sequence of events in $\AA_0$

\PROOF  Let $\bar\AA_0$ be the set of all such limits.
Because a $\sigma$-algebra can be defined as 
a Boolean algebra of sets 
which is 
complete in the sense that it is
closed under forming the union of an increasing sequence of sets,\footnote{$^\star$}
{Increasing means that $A^1 \subseteq A^2 \subseteq A^3 \cdots$.}
and because the $\sigma$-algebra generated 
by any family $\Buchstabe{F}$ of events 
is by definition the smallest $\sigma$-algebra that includes $\Buchstabe{F}$, 
it suffices to prove that 
$\bar\AA_0$ 
is closed under Boolean addition and multiplication,
and that forming the union of an increasing sequence members of $\bar\AA_0$ does not lead out of $\bar\AA_0$ either.
Since closure under the Boolean operations is the content of 
the 
preceding lemma,
we only need to demonstrate closure under nested countable union.  
To that end,
let $A=\bigcup\limits_j A^j$ be the union of an increasing sequence of events $A^j\in\bar\AA_0$,
each of which is the limit of a sequence of events $A^j_k$ in $\AA_0$.
It is a basic\footnote{$^\dagger$}
{Basic but quite simple to prove from the axioms for a measure [4].}  
%
result of measure theory (sometimes called ``continuity'')
that in this situation, $\mu(A\less A^j)\to0$. 
But because
$A^j\subseteq{A}$, 
$A+A^j = A\less A^j$, 
and we have 
$d(A^j,A)=\mu(A+A^j)=\mu(A\less A^j)\to0$.
Now choose $\eps>0$ and find an $A^j$ such that $d(A^j,A)< \eps/2$,
finding next an index $k$ such that $d(A^j_k,A^j)< \eps/2$.
Together, these imply that
$d(A^j_k,A)\le d(A^j_k,A^j)+d(A^j,A)\le \eps/2+\eps/2=\eps$, whence
$\AA_0$ contains events arbitrarily close to $A$, as required.

The proof of our zero-one law will rest on the previous lemma together
with the following one.

\LEMMA 7.  If events $A$ and $B$ are limits of sequences of events $A_j$
and $B_j$ respectively, and if for each index $j$, $A_j$ is stochastically
independent of $B_j$, then $A$ and $B$ are also stochastically independent.

\PROOF  By definition, stochastic independence of $A$ and $B$  signifies that
$\mu(AB)=\mu(A)\,\mu(B)$, which accordingly is what we want to prove.
But by hypothesis, we have
$\mu(A_jB_j)=\mu(A_j)\,\mu(B_j)$.
Appealing now to an earlier lemma,
we
can conclude from $A_j\to A$ that 
$\mu(A_j)\to\mu(A)$ 
and similarly 
$\mu(B_j)\to\mu(B)$, 
whence $\mu(A_j)\mu(B_j)\to\mu(A)\mu(B)$.
On the other hand, 
$A_jB_j \to AB$,
whence
$\mu(A_jB_j) \to \mu(AB)$,
completing the proof.

\section{A zero-one law and its proof}       
Let us say that an event $A\in\AA$ is {\it\/deterministic\/} if its
probability $\mu(A)$ is either 0 or 1, but nothing in between.  One also
says that $A$ obeys a ``zero-one law''.  If $A$ is a deterministic
event, then either it or its complement, $1+A$, is forbidden. In the
jargon of probability theory, an event forbidden in this way ``almost surely
will not happen'', while its complement ``almost surely will''.

Consider now some event $A$, 
let $G$ be the Poincar{\'e} group,
and let $g\in{G}$ act on $A$ by 
acting on the individual sprinklings $\omega$ that comprise it:
$gA=\SetOf{g\omega}{\omega\in A}$.  
By the {\it\/invariance group\/} of $A$ 
we mean the subset $H$ of $G$ 
whose elements leave $A$ unchanged.


\THEOREM  If the invariance group of an event $A$ contains at least one
non-zero spacetime translation then $A$ is a deterministic event
with respect to the Poisson process in $\Mink^n$.
\phantom{mmmmmmmmmmmmmmmmmmmmmmmmmmmmmmmmmmmmmmmmm}

\PROOF  Observe to begin with that if the invariance group $H$ contains
the translation $T$, it automatically contains all powers of $T$; 
it therefore contains arbitrarily large translations.  
It follows for any bounded spacetime region $K$ that $H$ contains a translation $T$ for which $K$ and $TK$ are disjoint.  
Now let $B$ be an event within the bounded region $K$, 
and choose $T{\in}H$ so that $K$ and $K'=TK$ are disjoint, 
and let $B'=TB$.
Since  
$B$ is an event within $K$ and  
$B'$ is an event within $K'$, 
and since $K$ is disjoint from $K'$,
$B$ will be stochastically independent of $B'$, 
this being 
a basic feature of  Poisson processes.

Now let
$A_k$ be a sequence of bounded events such that $A_k\to A$. 
Such a sequence exists by Lemma 6.
We have just seen that for each index $k$, there is a translation $T_k\in{H}$
such that $A_k$ and $A_k'=T_k A_k$ are stochastically independent.

Moreover, we claim 
(and this is the key to the proof)
that these translated events $A_k'$ {\it\/also\/}
converge to $A$.  To see why, recall first that by the definition of $H$,
the event $A$ is not altered by any of the $T_k$, i.e. $T_k A = A$. 
Then since the Poisson-process measure $\mu$ is itself translationally invariant, 
we have
  $d(A_k',A) = d(T_k A_k, A) = d(T_k A_k, T_k A) = d(A_k, A) \to 0$.
as claimed.

We now have two convergent sequences of events
whose individual terms are stochastically independent.  According to
Lemma 7, this entails that the limit-events are also
stochastically independent.  
But we just proved
that these limit-events are both equal to $A$, whence $A$ is independent
of itself!  
As an equation, this says that $\mu(AA)=\mu(A)\mu(A)$, or
$\mu(A)=\mu(A)^2$, since of course $AA=A$.  The only solutions of this
equation being $\mu(A)=0$ or $\mu(A)=1$, the theorem is established.


\section{Can a sprinkling break Poincar{\'e} invariance?} 
The theorem just proven will let us demonstrate several results that
rule out in various cases that a sprinkling can break one of the
symmetries of $\Mink^n$.  When combined with the analogous results from
[2], we expect that all cases of physical interest will be
spoken for.  To make this plausible we now apply our theorem to some
prototypical examples.  

\subsection {A sprinkling cannot determine an orientation}  
As a first example let's ask whether a Poisson sprinkling can break one
of the reflection-invariances by favoring either a particular spatial or
temporal orientation, or a particular overall orientation.  The
reasoning being the same in all these cases, let's take for definiteness
the case of an overall orientation (which is preserved by CPT but not CP
or T).  The question is then, Can a sprinkling --- an individual
realization of the Poisson process --- determine (with non-zero
probability) a specific orientation $O$?

Of course only two orientations are possible,
say $O_1$ and $O_2$, so our question reduces to asking for the
probability $p$ that the sprinkling will favor $O_1$ over $O_2$.
By symmetry $p$ is also the probability that it will favor
$O_2$ over $O_1$.  For maximum generality, we also admit that it might
favor neither, so that $p$ might be strictly less than $1/2$.  We claim
in fact that $p=0$.  

To prove this consider the event $A$ that the realization (call it $\omega$)
favors $O_1$.  Since an orientation can be thought of as an equivalence
class of orthonormal tetrads (if $n=4$), and since an orientation is
something global, the tetrads are located nowhere in particular (or if
you like they are located everywhere).  The event $A$ is thus trivially
invariant under all translations.  (If $\omega$ determines $O$ and if
$T$ is any spacetime symmetry, then $T\omega$ must
determine $TO$, which as we just saw, is $O$ itself when $T$ is a translation.)

Our theorem then informs us that $A$ is a deterministic event, whence
either $p=0$ or $p=1$.  But since $p\le1/2$ in any case, 
the only consistent
possibility is that $p=0$, as claimed.
Thus, a sprinkling will almost surely leave the reflections unbroken.



One might wonder whether something would go wrong here if the
sprinkling determined {\it\/more\/} than just an orientation.  What if
it also determined a distinguished location in spacetime, for example?
In fact nothing would go wrong because we assumed nothing about what
else $\omega$ might be able to determine.  The event $A$ would still be
defined and would still be translation-invariant because it would gather
together all the $\omega$ which favor $O_1$ irrespective of which
location they might also favor.

On the other hand, the doubt we have just sought to dispel does point to
a perennially confusing ambiguity that lurks in a phrase like
``A sprinkling cannot break T-reversal''.
Is it saying that the particular isomorphism $t\to-t$ is 
(in some coordinate system)
a symmetry
(meaning in the present context that it belongs to the invariance group $H$)
or is it only saying that 
a sprinkling cannot prefer a direction of time?
The difference shows up famously in discussions of the standard
model of high energy physics, where people are wont to say that
time-reversal is broken but that the laws of physics introduce no arrow
of time because CPT is a symmetry that reverses any putative arrow.  
{\it\/ What our proofs in this paper establish directly is the second kind of statement}, 
{\it\/which only indirectly bears on the first.\/}

\subsection {A sprinkling cannot break translation-symmetry by determining a spacetime lattice}
In the orientation example we just treated, the tetrads acted as a kind
of order-parameter or Higgs field responsible for the (putative)
symmetry breaking.  We take it as an article of faith that this will
always be the case: if a sprinkling breaks a spacetime symmetry it will
be because one can deduce from it some geometrical object $X$ whose
invariance group $H$ is a proper subset of the full group $G$ of
symmetries.  (In the case of Minkowski spacetime, which is our main
interest, $G$ will be the Poincar{\'e} group including all of its
connected components.  In the case of Euclidean space, to which our
analysis also applies, $G$ will be the Euclidean group, etc.) 

In the present subsection, we ask whether a Poisson sprinkling can break
the translation symmetry of spacetime.  For this to happen, $X$ would
have to be for example a distinguished ``origin'' in spacetime,
resulting in a trivial $H$ of no residual symmetry.  But $X$ could also
be a rectangular lattice of spacetime points, resulting in an $H$
identifiable with the subgroup of translations that preserve the
lattice.  (This situation is familiar from crystallization, and ``crystal
group'' might be an apt name for $H$.  As this name suggests, the full
$H$ might include some rotations, etc, but we will ignore them here
since our concern in this example is just with translations.  Thus we
will for now limit $G$ just to the translations.)

Suppose now that some sprinkling $\omega$ determines the lattice $L$.
Reasoning as before from the overall $G$-invariance of the Poisson
{\it\/process\/}, we see that other sprinklings must be able to
determine other lattices, all of them equally probable.  The lattices
obtainable in this manner can, in the familiar way, be identified with
the elements of the coset space $G/H$ (topologically a torus).

Fix now a particular lattice $L_1$, and let $p$ be the probability that
$L_1$ will result from a sprinkling.  Or more correctly (since we don't
want $p$ to vanish trivially),
introduce a
small rectangular neighborhood $\tilde{L_1}$ of $L_1$ and let $A$ be the
event: ``The sprinkling $\omega$ determines a lattice $L$ belonging to
$\tilde{L_1}$''.  
If the neighborhood $\tilde{L_1}$ was chosen suitably, 
$A$ will be invariant under $H$, 
the invariance group of $L_1$, 
and we define $p=\mu(A)$.


The event $A$ is the analog of the event of the same name in the
orientation example, and from here onward, we can proceed exactly as
before. 
On one hand, since $H$ contains nontrivial translations, $A$ is
deterministic, thanks to our theorem.\footnote{$^\flat$}
{In the previous example the full strength of our theorem was not
 needed, because $H$ there included the entire translation group.}
On the other hand, $p=\mu(A)<1$
because there are other ``fuzzy-lattice events'' which are just as
probable as $A$ is with respect to our Poisson process.  Therefore $p=0$
is the only possibility,
and a sprinkling will almost surely leave the translations unbroken.

\REMARK  Exactly the same argument goes through for lattices $L$ in
Euclidean space.

\subsection {A sprinkling cannot prefer a timelike direction: two methods of proof}
%
This was the main theorem proven in [2] by a different method
that assumed only that the sprinkling process was invariant under
Lorentz transformations.  In this paper, we are assuming more
specifically that our sprinkling process is a Poisson process.  To what
extent this is a loss of generality is unclear, since at present 
there seems to be no known example 
of a sprinkling process that is
Poincar{\'e} invariant without actually being Poisson (barring the
trivial exception of a convex combination of Poisson processes of
different densities).

Let us compare and contrast the two methods of proof.  

Following the pattern established with the previous two examples,
suppose that a sprinkling $\omega$ could determine the timelike unit
vector $u$.  Let $G$ be the Poincar{\'e} group, as before, and let
$H{\subseteq}G$ be the subgroup that acts as the identity on $u$.
The quotient $G/H$ can then
be identified with the (two-sheeted) unit hyperboloid in $\Mink^n$.
%
Consider as before the sprinkling-induced correspondence $\omega{\to}u$
and express it as a partial function $F:\Omega\to G/H$ (it is partial
because its domain might not be all of $\Omega$).  Continuing to reason
as before, we learn that $F$ induces on $G/H$ a (subnormalized)
probability distribution $\nu$.  Because it must be invariant under $G$,
we know also that $\nu$ could only be a constant density on $G/H$.

At this point the two methods part ways.  The method of [2]
simply notices that unless $\nu=0$ its integral over all of $G/H$ would
be infinite, whereas in fact it cannot exceed unity (being
subnormalized). The only way out of this contradiction is that the
domain of $F$ is a measure-zero subset of $\Omega$.  The method of this
paper, on the other hand, reaches the same conclusion by introducing a
bounded subset $S$ of $G/H$ and observing that the event $A$ given by
``$F(\omega)\in S$'' is translation invariant since $u$ is a global
object, like the orientations in our first example.  Hence $A$ is
deterministic, and $\nu(S)=\mu(A)$ can only be 0 or 1, whence it must be
0 since it cannot be 1. 

How then do the two methods differ?  Both proceed from the same uniform
density $\nu$ on $G/H$, but they presuppose different things about $H$ and $G/H$.  
The first method lives off the fact that $G/H$ has an infinite volume.
The second lives off the fact that $H$ contains a nontrivial translation.
Thus, the  first method works when $H$ is ``sufficiently small'',
the second works when $H$ is ``sufficiently big''
(but not so big that $G/H$ fails to contain at least two points.  
In that case 
$H=G$ and there is no breaking at all.)

In the previous two
examples, the first method would not have worked because $G/H$ was
compact and hence of finite volume.  On the other hand the second method
would have trouble if the sprinkling were trying to break
translation-invariance completely 
by picking out a unique favored point or ``origin''; 
in that case $H$ would contain no translations.
We would conjecture that in all cases of interest at least one of the
two methods will work.  This would be true, for example, if $G/H$
necessarily had finite volume whenever $H$ failed to contain a
translation.

\subsection{A sprinkling cannot prefer a ``lattice'' of timelike directions}
%
As a last illustration of the second method, let us consider the
possibility that `$X$' is not a single timelike direction but an
infinite set of them which is invariant under a discrete subgroup of the
Lorentz group $G$.\footnote{$^\star$}
{To be mathematically impeccable, we should point out that $G$ here is
not literally a subgroup of the Poincar{\'e} group, but of its quotient
by the translations.  That is, $G$ doesn't act on spacetime itself,
which is strictly speaking an affine space, but rather on the
associated vector-space.}
It might seem surprising that such a subgroup exists at all, but many
instances are known.  One of the most interesting is comprised of the
set of Lorentz transformations that leave invariant the integer lattice
$\Integers^4$ in $\Mink^4$ [6] [7].  The elements of $X$ itself can
then be taken to be the unit vectors pointing from the origin to
the points of $L$.  Let us focus on this example.

It seems that there are general theorems of Algebraic Geometry which
imply in this case that 
orbit of such an $X$ under the action of the Lorentz group,
though not actually compact, 
has only a finite volume [8].
Our first method of proof would then not apply.  The second method does
apply however for the same reason it applied to a single timelike
direction, our $X$'s being by definition translation invariant.

\section{What does it all mean?}             
We don't have access to all of spacetime, and in any case we don't live
in $\Mink^4$.  What then is the physical relevance of theorems about
sprinklings of a flat spacetime?  Recall that the sprinkling of a
Lorentzian manifold $M$ has only a kinematical and not a dynamical
significance.  It is meant to provide a causal set typifying those that
could be the substructure of $M$.\footnote{$^\dagger$}
{Even this statement ignores that quantum spacetime is expected to be
 more like a ``superposition'' of causal sets than a single one.
 Moreover, we only expect a sprinkling to be a good model after a certain
 amount of coarse-graining, e.g. if at small scales the structure of
 spacetime were of Kaluza-Klein type.}
If in this paper we have taken $M$ to be literally $\Mink^4$, 
this is only an idealization
of some approximately flat region $R$ within the larger universe.
What we'd really like, then, is not only a global proof of Poincar{\'e}
invariance, but a quasilocal result that would quantify how much
anisotropy or inhomogeneity remains, depending on the size of $R$.
Our rigorously proven theorems are but a first step toward such an analysis.
(As usual there's a trade-off between beautiful theorems and applicability!)

In Euclidean space, such an analysis seems near at hand.  To each
spatial point $x$ we can associate the line that passes through it and
its nearest sprinkled neighbor.  Each such line breaks the rotation
symmetry {\it\/at that point\/} to $\Integers_2$, which is of course why
rotations cannot literally be a symmetry of a sprinkling but only so in
an average
sense.  It is equally clear, though, that these lines
fluctuate wildly in direction, so the anisotropy dies out rapidly with the
size of the region one considers.
Similarly, one would expect any localized inhomogeneities to wash out on
larger scales so that translation-invariance would return.

In Minkowski space something similar is plausibly true, but in
relation to the Lorentz subgroup of the Poincar{\'e} group,
there's a complication; 
both the size {\it\/and\/} the shape of the region $R$ are important.
Nevertheless we would still expect to get a rapidly fluctuating array of
lines that are, in the natural rest-frame of the region,\footnote{$^\flat$}
{What is the ``frame of the region''?  Well, find two points $x$, $y$ in
$R$ such that the order-interval $I(x,y)$ has the biggest volume
possible.  The line through $x$ and $y$ then defines the rest-frame in
question.  Some such prescription ought to be adequate in most cases.}
nearly null,
and so the breaking would again die out rapidly as $R$ grew.  Only now
in a finite region we won't restore all of the Lorentz group, but only
those boosts that are small enough for $R$ to accommodate.  This
``boundary effect'' (or ``shape effect'') has no analog in the Euclidean
case, but otherwise the two situations seem quite similar.

Beyond these kinematic questions of global theorems vs. quasilocal
applicability, what we ultimately care about are consequences for the
dynamics.  Would a massless scalar field living on a Poisson sprinkling
propagate via a modified dispersion relation, as has been suggested for
discrete structures?  
The answer depends obviously on how the dynamics is formulated, so it is
impossible to answer categorically.  
But our theorems are significant precisely because 
they indicate that the answer will be ``No''. 
(We ignore here the possibility of dynamical spontaneous symmetry breakings
which have nothing to do with kinematical discreteness.)

Which doesn't mean there might not be {\it\/other\/} ``dispersive'' or
diffusive effects consistent with all the spacetime symmetries.  We hope
that there are, because they would be highly constrained by the symmetry,
and would potentially provide phenomenological evidence of discreteness!
[1] [9]
Indeed, such effects, although not yet seen experimentally or
observationally, have already begun to be studied in extant theories
that describe the dynamics of particles and/or fields on a background
causal set.
(For examples of such theories, see [10])

But even these reflexions are not the end of the story.  Beyond dynamics
on a fixed, background causal set, we need ultimately to understand the
effects of the causal set itself being dynamical (i.e. of quantum
gravity).  Our theorems here are merely a first indication of how things
are likely to turn out.






\bigskip
\noindent
We thank Adrian Kent for discussions which inspired us to ask whether
zero-one laws might be provable for Poisson processes, and if so whether
they could help us to extend the conclusions of [2] about 
symmetry-breaking (or its absence) by sprinklings.
This research was supported in part by NSERC through grant RGPIN-418709-2012.
This research was supported in part by Perimeter Institute for
Theoretical Physics. Research at Perimeter Institute is supported
by the Government of Canada through Industry Canada and by the
Province of Ontario through the Ministry of Economic Development
and Innovation.  
FD is supported in part by STFC grant ST/P000762/1 and APEX grant APX${\backslash}$R1${\backslash}$180098.

\section{Appendix: reply to Adrian Kent [11]}            
In a recent paper [11], Adrian Kent has disputed our
interpretation of the theorems proven in [2].  As far as we can
see, he puts forward three main criticisms, and we take this opportunity
to explain why we think they are unfounded.  We hope also, that our
comments will help bring into focus the conceptual background to both the
work in [2] and its extension here.

Kent's primary complaint seems to be that attention should fall on what he
calls ``sprinklable sets'' instead of sprinklings, where a sprinklable
set is an isometry equivalence class of sprinklings. 
This amounts to treating Poincar{\'e} symmetries as if they were merely
gauge, contrary to the way most physicists understand them.
(We follow here the widely used terminology that draws a distinction
 between ``gauge transformations'' that, like coordinate
 transformations, merely alter the description without affecting
 physical reality, and ``symmetries'' which effect genuine physical
 changes.  It is, for example, because one treats translations as
 symmetries that it is meaningful to speak of the energy-momentum
 vector of a system.)
We believe that the majority viewpoint is in this case the appropriate
one.  As highlighted earlier, we don't live in $\Mink^4$ but in a cosmos
that is highly curved on large scales and near to black holes, etc.  In
such a universe a flat spacetime can only be an idealization of a nearly
flat local region $R$.  But as soon as you remember that all such
regions exist within an enveloping spacetime, you realize
[12] that local translations, rotations, and
Lorentz-boosts are in the larger context {\it\/not\/} pure gauge,
because they move a subsystem around relative to its environment.  They
are rather real physical changes idealized as what one might term
``partial gauge transformations'';\footnote{$^\star$}
{By {\it\/partial gauge transformation\/} we mean an operation which is
 locally indistinguishable from a gauge transformation but which only
 acts nontrivially on a subsystem or region while leaving the
 surroundings unchanged.  Most if not all symmetries can be understood as
 partial gauge transformations.  See for example the brief discussion of
 this concept (though not under this name) in \S1 of [13]}
and one really ought to think of $\Mink^4$ as being referred to an
``external frame'' --- a laboratory, the fixed stars, etc.  (If the
whole of spacetime really were $\Mink^4$, one might have to rethink the
status of the Poincar{\'e} group, but obviously that is not the case.)
Thus sprinklings and not sprinklable sets are the appropriate objects of study. 

Having replaced sprinklings by sprinklable sets, Kent then argues, if we
understand him, that the zero-one law that holds for propositions about
sprinklable sets is a bad thing because it means in some sense that
one cannot say anything interesting about a sprinklable set created by a
Poisson process.
Of course, this criticism cannot be sustained if, as we have just
argued, it is sprinklings and not sprinklable sets that are physically
relevant.  
But instead of just stopping with this comment, perhaps we should add that (as
explained by Kent himself under the heading ``A Lacuna in the BHS
Theorem'') a question like ``Does the sprinkling determine a timelike
direction?'', still makes sense as a question about sprinklable sets.
The corresponding event in the sample-space $\Omega_S$ of sprinklable
sets is simply the union of all the events in $\AA$ that belong to
specific timelike directions; and it still has measure zero.  (See
[14] for how $\Omega_S$ is related to $\Omega$.)
Since this question and others like it hold the keys to
deciding whether a sprinkling can break a spacetime symmetry, we cannot
agree that the $\sigma$-algebra of $\Omega_S$ is too sparse to contain
events of physical interest, even if one chooses to study it instead of $\Omega$.

But independent of ``sprinkling vs. sprinklable'', could it be that
something else is behind the criticism?  There are hints in [11]
that 
one
is thinking of the Poisson
process as a kind of dynamical theory of causal sets.
If one were to think of it in this way, then one
might feel uncomfortable that every event
in this theory would be deterministic.
%
%
For some purposes that might be an interesting observation, 
but it is in any case not relevant to causal-set dynamics. 
As described in the previous section,
sprinklings within causal set theory
play only the kinematical role of 
helping to define the relationship
between a causal set and the corresponding spacetime continuum.
Dynamical laws (``laws of motion'')
meant for causal sets can presuppose no background
spacetime, 
and
are envisioned as defining a stochastic process of growth which, as it
were, builds up an evolving causal set element by element.

\REMARK Suppose that in some context one actually did want to interpret
the Poisson process as a discrete dynamics for Minkowski
spacetime. 
There is only one $\Mink^4$-geometry, and since every question you can
ask about its structure thereby has a unique yes-or-no answer, would not
a zero-one law for such questions be exactly what you would want?  It would 
suggest 
that your dynamics had
reproduced $\Mink^4$ as well as it could consistent with discreteness.

%
 
Kent's third criticism seems to be that reference [2] proved
the wrong thing, or at least failed to prove some things it needed to
prove.  In effect he has brought forward a new requirement that anyone
claiming to establish Poincar{\'e} invariance needs to satisfy, which he
states as follows.  ``One needs to show that, given any data that leave
some continuous subgroup of the Lorentz group as a symmetry in the
continuous case, there is no mathematical construction that breaks this
symmetry in the discrete case.''

To see what this means, consider for simplicity the Euclidean question
whether a sprinkling can prefer a spatial direction, thereby breaking
isotropy.  This was a question that could not be answered in
[2], but which we have answered in the negative in the present
paper.

Now consider the different question whether a sprinkling could determine
a spatial direction if one provided in addition a marked spatial point
or ``origin''.  As pointed out in [2], the answer to this
question is ``yes''.
Does this constitute a breaking of isotropy?  Kent thinks it does,
whereas we think it does not, because the required extra information is
in reality absent.\footnote{$^\dagger$}
{In the Lorentzian example considered in [11], the extra
 information is a timelike direction, but one is still asking about
 spatial rotations.}
We therefore disagree that there is some kind of ``lacuna'' in the
theorems of [2] or this paper.  For us the most pertinent
questions are the {\it\/intrinsic\/} ones, that ask whether a sprinkling
{\it\/in and of itself\/} can break a symmetry.

The above is of course not meant to claim that the theorems in
[2] settled every question one might want to ask.  On the
contrary, our concern in this paper has been to complement those
theorems by analyzing a larger class of symmetry-breaking scenarios than was
possible with the tools of [2] alone.  And beyond that loom the
whole series of questions adumbrated in the previous section.

It is connection with the latter questions that Kent's ``extra
information'' might become relevant.  He invokes for example a particle
moving through a medium of sprinkled points (in $\Mink^4$, but let's
stay Euclidean for convenience).  The particle itself ``marks a point'',
and so it can in fact see some anisotropy.  It will then swerve from a
straight line, and this effect could be noticed.  Very good!  This is
precisely the type of effect one expects from discreteness.  But what's
important is the inference that  
--- precisely because isotropy is intrinsically {\it\/preserved\/} --- 
the diffusion equation describing these swerves will
be rotationally invariant.  Just such an equation, in its Lorentzian guise,
was brought forth in [1] as a possible phenomenological
manifestation of an underlying causal set.  The extrinsic information
provided microscopically does something observable, but in a manner that
respects the intrinsic global symmetry.

\REMARK Apropos of Kent's remarks on local Lorentz invariance, we
have noticed that certain passages in [2] could lead readers 
to interpret that ambiguous phrase in a manner less like what
it would mean in the context of this paper, and more like what it means
in connection with fields of orthonormal tetrads.
If so, we hope that the reflections in the previous section concerning
what one might call ``local Poincar{\'e}-invariance'' (which, be it
noted, includes translations) will have made it clear that 
the words local or quasilocal are in the present context not meant
to point to any extrinsically given location or marked point in
spacetime; they are meant rather to evoke the kind of approximately flat
region $R$ expounded on above under the heading ``What does it all
mean?''

\ReferencesBegin                             

\ref [1] Fay Dowker, Joe Henson and Rafael D.~Sorkin, ``Quantum Gravity Phenomenology, Lorentz Invariance and Discreteness'',
\journaldata {Modern Physics Letters~A} {19} {1829-1840} {2004} \lbr
\eprint{gr-qc/0311055} \lbr
\eprint{http://www.pitp.ca/personal/rsorkin/some.papers/115.swerves.pdf}

\ref [2] Luca Bombelli, Joe Henson and Rafael D. Sorkin, ``Discreteness without symmetry breaking: a theorem''
 \journaldata {Modern Physics Letters} {A24} {2579-2587} {2009} \lbr
 \eprint{gr-qc/0605006} \lbr
 \eprint{http://www.pitp.ca/personal/rsorkin/some.papers/121.sprinkling.pdf}

\ref [3] G{\"u}nter Last and Mathew Penrose, {\it\/Lectures on the Poisson process, volume 7\/} (Cambridge University Press, 2017)

\ref [4] A.N.~Kolmogorov and S.V.~Fomin, {\it Measure, Lebesgue Integrals, and Hilbert Space},
 translated by Na\-tascha Artin Brunswick and Alan Jeffrey (Academic Press 1961)

\ref [5] J.F.C. Kingman, {\it Poisson Processes}, Oxford Studies in Probability, 3 (Clarendon Press, Oxford 1993)

\ref [6] Alfred Schild, ``Discrete space-time and integral Lorentz transformations'' \journaldata{Canadian J. Math}{1}{29-47}{1949}

\ref [7] Mehdi Saravani and Siavash Aslanbeigi, ``On the Causal Set-Continuum Correspondence'' \eprint{arXiv:1403.6429 [hep-th]}

\ref [8] Avner ash personal communcation.

\ref [9] 
Lydia Philpott, Fay Dowker, and Rafael D.~Sorkin, ``Energy-momentum diffusion from spacetime discreteness'', 
\journaldata {Phys. Rev. D} {79} {124047} {2009}
\eprint{arxiv:0810.5591 [gr-qc]},
\eprint{http://www.pitp.ca/personal/rsorkin/some.papers/130.swerves.II.pdf}

\ref [10] 
Rafael D.~Sorkin, ``Scalar Field Theory on a Causal Set in Histories form'' 
 \journaldata{Journal of Physics: Conf. Ser.}{306}{012017}{2011},
 \arxiv{1107.0698},
 \eprint{http://www.pitp.ca/personal/rsorkin/some.papers/142.causet.dcf.pdf};
\sepref
Edmund Dable-Heath, Christopher J. Fewster, Kasia Rejzner, Nick Woods, 
``Algebraic Classical and Quantum Field Theory on Causal Sets'', \eprint{arXiv:1908.01973} [math-ph];
\sepref
Mehdi Saravani, Rafael D.~Sorkin, and Yasaman K.~Yazdi, ``Spacetime Entanglement Entropy in 1+1 Dimensions''
  \journaldata{Class. Quantum Grav.}{31}{214006}{2014} 
   available online at http://stacks.iop.org/0264-9381/31/214006
  \arxiv{1205.2953}
  \eprint{http://www.pitp.ca/personal/rsorkin/some.papers/148.pdf};
\sepref
Rafael D.~Sorkin and Yasaman K.~Yazdi, ``Entanglement Entropy in Causal Set Theory'' 
 \journaldata{Class. Quantum Grav.} {35} {074004} {2018}, 
 \arxiv{1611.10281}; 
\sepref
%
%
Nosiphiwo Zwane, ``Cosmological Tests of Causal Set Phenomenology'' (doctoral thesis), Section~3\lbr
\eprint{uwspace.uwaterloo.ca/bitstream/handle/10012/12414/Zwane\_Nosiphiwo.pdf};
\sepref
Fay Dowker, Joe Henson and Rafael D.~Sorkin, ``Discreteness and the Direct Transmission of Light from Distant Sources'',
 \journaldata {Phys. Rev. D} {82} {104048} {2010}\lbr
 \arxiv{1009.3058};
\sepref
Rafael D. Sorkin, ``Does Locality Fail at Intermediate Length-Scales?''
 in  {\it Approaches to Quantum Gravity -- Towards a new understanding of space and time}, edited by Daniele Oriti 
 (Cambridge University Press 2009) pages 26-43, \lbr
 \eprint{gr-qc/0703099},\lbr  
 \eprint{http://www.pitp.ca/personal/rsorkin/some.papers/122.nonlocality.pdf};
\sepref
Mehdi Saravani, Rafael D.~Sorkin, and Yasaman K.~Yazdi,
``Spacetime Entanglement Entropy in 1+1 Dimensions''
  \journaldata{Class. Quantum Grav.}{31}{214006}{2014} 
   available online at http://stacks.iop.org/0264-9381/31/214006
  \arxiv{1205.2953}
  \eprint{http://www.pitp.ca/personal/rsorkin/some.papers/148.pdf}.

\ref [11] Adrian Kent, ``Are There Testable Discrete Poincar{\'e} Invariant Physical Theories?'',
\journaldata{Class. Quantum Grav.}{35}{195001}{2018}

\ref [12] Henri Poincar{\'e}, {\it\/Mathematics and Science: Last Essays\/}, 
translated from the French by John W. Bolduc
(New York, Dover 1963),
pp.~20-21

\ref [13] John L.~Friedman and Rafael D.~Sorkin, ``Statistics of Yang-Mills Solitons'', \journaldata{Comm. Math. Phys.}{89}{501-521}{1983}, see \S1.

\ref [14] Graham Brightwell, Fay Dowker, Raquel S.~Garc{\'\i}a, Joe Henson and Rafael D.~Sorkin, ``{$\,$}`Observables' in Causal Set Cosmology'',
\journaldata {Phys. Rev.~D} {67} {084031} {2003} \lbr
\eprint{gr-qc/0210061} \lbr
\eprint{http://www.pitp.ca/personal/rsorkin/some.papers/110.obs.pdf}


\end


(prog1 'now-outlining
  (Outline* 
     "\f"                   1
      "
      "
      "
      "
      "\\Abstract"          1
      "\\section"           1
      "\\subsection"        2
      "\\appendix"          1       ; still needed?
      "\\ReferencesBegin"   1
      "
      "\\ref "              2
      "\\end